\documentclass[preprint]{aastex}

\newcommand\simlt{\lower.5ex\hbox{$\; \buildrel < \over \sim \;$}}
\newcommand\simgt{\lower.5ex\hbox{$\; \buildrel > \over \sim \;$}}

\slugcomment{To appear in the {\it Astrophysical Journal Letters},
DATE}

\begin{document}
\title{A calorimetric test for Kerr black holes
       in gamma-ray bursts}
\author{Maurice H.P.M. van Putten}
\affil{MIT, Cambridge, MA 02139-4307}

\begin{abstract}
  Cosmological gamma-ray bursts are potentially associated with
  systems harboring a rotating black hole. Calorimetry on yet 
  ``unseen" emissions in gravitational waves from a surrounding 
  torus may provide a measure for its rotational energy.  
  We introduce a compactness parameter $\alpha=2\pi \int fdE$ 
  for the fluence $E$ and frequency $f$ in graviational waves.
  For black hole-torus systems, $\alpha\simeq 2\pi Ef\simeq 
  0.005-0.05$ may exceed the upper bound $\alpha^*\simeq0.005$ for rapidly 
  rotating neutron stars. A duration in gravitational 
  radiation consistent with the redshift corrected durations of the
  BATSE catalogue defines an association with long GRBs. 
  This promises a new test for Kerr black holes as objects in Nature.
\end{abstract}


\section{Introduction}
  
  Gravitation is a vital and universal force in Nature.
  It ultimately leads to black holes as fundamental 
  particles according to the Einstein theory of 
  general relativity. These objects could be the source of
  cosmological gamma-ray bursts (GRBs) - the most enigmatic
  events in the Universe. Their emissions are characteristically
  non-thermal in the few hundred keV range with a bi-modal
  distribution in durations, of short bursts around 0.3s and
  long bursts around 30s\citep{kou93}. These  probably
  derive from dissipation of the kinetic energy in ultrarelativistic 
  baryon poor jets (see \cite{pir98} for a review). 
  Redshift determinations from the
  X-ray afterglows to long bursts indicate a cosmological origin,
  consistent with their isotropic distribution in the sky
  and $<V/V_{max}>$ distinctly less than the Euclidean value of 1/2
  \citep{sch99}. They are probably associated with the formation of 
  young massive stars \citep{pac98,blo00} and, hence, should be most
  frequent at a redshift of about two. Remnants of GRBs are potentially
  found in some of the soft X-ray transients \citep{bro00}, with 
  notable candidates GRO J1655-40 \citep{isr99} and V4645 \citep{oro01}.
  
  The inner engine producing the GRBs should be energetic and compact.
  Angular momentum forms a canonical energy reservoir, and GRB inner
  engines are probably no exception. This is consistent with breaking
  of spherical symmetry \citep{woo93}. Hypernovae, then, follow up on 
  this idea by postulating massive young stars in binaries as their 
  progenitors \citep{pac98}. These proposals have in common the formation of a
  black hole, which should be rotating \citep{bar70} by conservation of angular 
  momentum from the progenitor star and its binary motion. Black holes 
  formed in prompt collapse, furthermore, are subject to a minimum
  mass of the order of $10M_\odot$ \citep{mvp01a} consistent with the
  observed masses of $3-14M_\odot$ in SXTs. 
  We note that high angular momentum is also present in coalescing
  black hole-neutron star systems.
  Encysting rotating black holes as the baryon-free energy source
  in GRBs is perhaps tantamount to understanding how do stars shine.

According to the Kerr solution of rotating black holes
\citep{ker63}, the horizon surface encloses baryonic 
matter collapsed under its own gravitational forces, 
while leaving rotational energy accessible by the Rayleigh 
criterion. The latter is a consequence of the first law of 
thermodynamics \citep{bar73,haw75,haw76},
\begin{eqnarray}
\delta M=\Omega_H\delta J_H+T_H\delta S_H,
\end{eqnarray}
where $\Omega_H\le 1/2M$ denotes the angular velocity of the 
black hole of mass $M$, $J_H$ its angular momentum, $T_H$ and 
$S_H$ its surface temperature and entropy, respectively.
The specific angular momentum $\delta J_H/\delta M$
of a radiated particle, therefore, is at least twice that 
of the black hole itself. In isolation, rotating black 
holes are effectively non-radiating due to angular 
momentum barriers \citep{unr74,haw75,teu73,pre73,teu74}.
However, a surrounding torus may trigger a rotating
black hole into violent activity (see \cite{mvp01b} 
for a review) -- perhaps epitomized in GRBs.

Identifying Kerr black holes as the active nucleus 
in GBRs may derive from its three defining properties: 
\\
{\em 1. a compact horizon surface}, which causally separates 
collapsed matter from its progenitor surroundings. The
black hole is expected to represent the final outcome
of stellar evolution, formed in accretion
induced collapse of a neutron star, white dwarf or, 
alternatively, in prompt collapse. This may produce
black holes of intermediate spin in the former \citep{bro99,bro00}, 
and black holes with rapid spin and above a certain mass threshold 
in the latter \citep{mvp01a}.
\\
{\em 2. frame-dragging of space-time}
in violation with Mach's principle, in the sense of
a correspondence of zero angular velocity relative to 
the distant stars with zero angular momentum. 
A rotating black hole introduces
differential rotation in space-time described by frame-dragging $-\beta$,
equal to the angular velocity $\Omega_H$ of the black hole
on the horizon and zero at infinity. Upon reaching the north pole,
a ballarina rotating with angular velocity $-\beta$ relative to the 
distant stars will find her arms naturally pointed down, while one standing 
still will find her arms spread out.
\\
{\em 3. a compact energy reservoir} of up to one-third of 
the total mass of the black hole. Rotation of a 
Kerr black hole may be parametrized by $\sin\lambda=a/M$,
where $a$ denotes the specific angular momentum $J_H/M$
for an angular momentum $J_H$ and mass $M$. This gives
$\Omega_H=\tan(\lambda/2)/2M$ and 
$E_{rot}= 2M\sin^2(\lambda/4)$ for the angular velocity
and rotational energy of a Kerr black hole, respectively.
Thus, the rotational energy may reach about 29\% of the
total mass of the black hole. This has no counterpart in
baryonic matter. About one-half of the rotational 
energy is stored in the upper ten percent of the
angular velocity of a maximally spinning black hole.

The first two properties -- a compact horizon and frame-dragging
-- are potentially the driving agency behind the formation of a 
macroscopic gap with ultrarelativistic baryon poor outflows along open magnetic 
field-lines \citep{mvp01b,mvp01e}, serving as the input to GRB/afterglows.
These field-lines find their support in an equilibrium magnetic moment
of the black hole in its lowest energy state as a consequence of the
no-hair theorem, when surrounded by a torus magnetosphere. The latter 
is expected to be supported by a disk or torus, from fallback matter 
in failed supernovae/hypernovae \citep{woo93,pac98} or debris of 
a neutron star upon tidal break-up \citep{pac91}.

Establishing the third property requires calorimetry on the 
output from GRBs in all energy channels. For long bursts from
rapidly spinning black holes \citep{mvp01a}, discussed here, 
we expect a high incidence of the
black hole luminosity onto the inner face of the surrounding disk 
or torus by equivalence in poloidal topology to pulsar magnetospheres
\citep{mvp99}.
Firstly, this output into the disk or torus may be processed into several channels,
powering gravitational waves, Poynting-flux dominated winds and, when
sufficiently hot, neutrino emissions. 
Indeed, evidence for appreciable energy output from the spin of the black hole
into the disk may already be found in chemical depositions onto the companion 
star of GRO J1655-40 and V4641Sgr, in the hypernovae proposed by \cite{bro00}.
Secondly, this output may be
representative for the total output of the black hole, being at least
two order of magnitudes more luminous \citep{mvp01c} than the
true output of $3-5\times 10^{50}$ergs in GRB/afterglows 
\citep{fra01,pan01,pir01}.
Thus, GRBs could be merely the tip of the iceberg, being accompanied by 
as yet ``unseen" emissions such as in gravitational waves, whose fluence
may reach about 1\% of the mass of the black hole\citep{mvp01d}.
Therefore,
the upcoming Laser Interferometric Gravitational Wave Observatory
LIGO \citep{abr92} and their French-Italian counter part VIRGO \citep{bra92}
promise a new avenue for calorimetry on the energy reservoir in GRBs, 
and possibly that of rotating black holes.

The association between the fluence in gravitational waves from a torus 
and the spin-energy of a central black hole is described in \S2. A compactness
measure for the observed fluence and frequency in gravitational waves is
proposed in \S3, which may serve to discriminate black hole-torus systems 
from rapidly rotating neutron stars.  We shall use geometrical units,
expressing energy $E$ in terms of the equivalent gravitational radius 
$EG/c^2=1.5\times 10^5\mbox{cm}(E/M_\odot c^2)$,
where $G$ denotes Newton's constant and $c$ the velocity of light,
and frequency $f$ in terms of $f/c$. For example, in these units 
luminosity is dimensionless, given as a rate of change of energy [cm]
per unit of time [cm].

\section{Proposed source of gravitational radiation from a torus around a black hole}

Long/short GRBs may be associated with a state of suspended-/hyperaccretion 
onto the black hole \citep{mvp01a}. A suspended-accretion state lasts as long
as the black hole spins rapidly, whereas the disk around a slowly rotating
black hole evolves by 
magnetic regulated hyperaccretion towards a finite-time singularity.
A bi-modal distribution of duration occurs, when the ratio of disk mass to
torus mass is small. The suspended accretion state operates by equivalence 
in poloidal topology to pulsar magnetospheres \citep{mvp99,bro01}: the inner
face of the torus receives energy and angular momentum from the black hole as 
does a pulsar when infinity wraps around it \citep{mvp01e}.
Gravitational radiation may be produced in black hole-torus systems
as the torus develops non-axisymmetric instabilities. With forementioned
association to GRBs, this features several aspects which suggest considering 
long GRBs as LIGO/VIRGO sources:
the torus is strongly coupled to the spin-energy of the black hole;
lumpiness in the torus will produce gravitational radiation at
twice the Keplerian angular frequency, i.e., in the range of
1-2kHz; gravitational radiation should dominate
over emissions in radio waves; the true rate of GRBs should be
frequent given their beaming factor of a few hundred
\citep{fra01,pan01,pir01}.
Powered by the spin-energy of the black hole, these emissions
are different from radiation produced during spiral in of neutron-neutron 
star binaries \citep{nar92,koc93}
or fragmentation in collapse towards supernovae \citep{bon95}. 

Non-axisymmetries in the torus are expected from dynamical and,
potentially, radiative instabilities. This can be described by
a lumpiness $m$, which introduces a luminosity according to the
Peters \& Mathews' formula for quadrupole radiation from two
point masses \citep{pet63}
\begin{eqnarray}
{\cal L}=\frac{32}{5}\left(\omega {\cal M}\right)^{10/3} F(e)\simeq
           \frac{32}{5}\left({M}/{R}\right)^5\left({m}/{M}\right)^2
\label{EQN_L1}
\end{eqnarray}
where $\omega\simeq M^{1/2}/R^{3/2}$ denotes the orbital frequency with 
separation $R$, 
${\cal M}=(M_1M_2)^{3/5}/(M_1+M_2)^{1/5}\simeq M(m/M)^{5/3}$
the chirp mass and $F(e)$ a factor of order unity as a function of the ellipticity $e$.
The torus may be thick, consistent with the recent indication that long GRBs
may be standard \citep{mvp01c}, which is generally susceptible to the Papaloizou-Pringle
instability \citep{pap85}. A massive torus, reaching an appreciable fraction of the
mass of the black hole, would further be unstable to self-gravity \citep{woo94}.
Non-axisymmetries might also be promoted by gravitational radiation, as in
the Chandrasekhar-Friedman-Schutz instability, or by its action preferentially 
on lumps of matter on inner orbits.
Gravitational wave-emissions thus produced could be quasi-periodic (QPO), perhaps
reminiscent of QPOs in X-ray binaries and possibly also
related to general relatistic effects in
orbital motions \citep{ste00}.  

We find that about one-third of the black hole-luminosity
$L_H$ (most of which is incident onto the torus) is re-radiated 
into gravitational waves \citep{mvp01d}:
\begin{eqnarray}
L_{gw}\simeq L_H/3
\label{EQN_LH}
\end{eqnarray}
(the fraction on the right hand-side may range from 1/4 to 1/2). 
For a maximally spinning black hole, therefore, the
fluence in gravitational radiation may reach about 
one-third the rotational energy of the black hole, times 
the efficiency factor given by the ratio of the angular
velocity $\Omega_T$ of the torus to that of the central rotator.  
For a torus of radius $R$ around a maximally spinning Kerr
black hole of mass $M$, the latter is given by \citep{sha83}
\begin{eqnarray}
{\Omega_T}/{\Omega_H}={2}/[{(R/M)^{3/2}+1}].
\label{EQN_EH}
\end{eqnarray} 
Hence, by (\ref{EQN_L1}-\ref{EQN_EH}), the characteristic mass $m$ 
of lumpiness in the torus is required to be 
\begin{eqnarray}
        m\simeq 0.15\% M (R/3M)^{7/4}
\end{eqnarray}
for a canonical burst duration of 15s (of long bursts). 
This estimate is consistent with condition that the perturbation $m$ is much 
less than the mass $M_T$ of the torus, while $M_T<< M$ 
for there to be a bi-modal distribution.
Note that the efficiency (\ref{EQN_EH}) will be less than 50\% for a torus radius $R>2M$, 
showing that generally most of the rotational energy is dissipated in the horizon of 
the black hole. Since 90\% of the rotational energy of a maximally spinning
black hole is contained in the upper 10\% of its angular velocity, a black
hole-torus system is not expected to evolve significantly during most of its
gravitational wave emissions. 

\section{Compactness parameter for bursts of gravitational radiation}

We propose a compactness parameter for gravitational wave-emissions 
by integration of the frequency $f$ against fluence: 
\begin{eqnarray}
\alpha= 2\pi \int_0^E f dE
\label{EQN_A}
\end{eqnarray}
for a net fluence $E$. An equivalent expression is the integration
of the luminosity ${\cal L}$ against the number of periods $n$, i.e.,
$\alpha= 2\pi\int_0^T {\cal L}(t)dn(t)$
over the duration $T$ of the burst. The measure (\ref{EQN_A}) is 
reminiscent of the ratio of the rotational energy 
$E_{rot}$ to the length scale $M$ of black hole. Indeed,
$\alpha\simeq 2\pi E/L$ for a system of size $L$ close to its 
Schwarzschild radius, 
whose fundamental angular frequency is of order $2\pi/L$. 
 In geometric units, 
$\alpha$ is dimensionless.

For a black hole-torus system, (\ref{EQN_A}) becomes
\begin{eqnarray}
\alpha\simeq 2\pi Ef
\end{eqnarray}
upon neglecting secular evolution while the black hole continues 
to spin rapidly. For a maximally spinning black hole, 
(\ref{EQN_LH}) shows that $E$ may reach about 
one-third the rotational energy of the black hole, times 
the efficiency factor (\ref{EQN_EH}). Hence, we have
\begin{eqnarray}
\alpha\simeq\frac{4}{9}[(R/M)^{3/2}+1]^{-2}
\simeq 0.005-0.05,
\label{EQN_T1}
\end{eqnarray}
where $R=(2-4)\times M$ denotes a fiducial range of torus radii, 
corresponding formally to $f=1.1-2.5$kHz for $M=7M_\odot$.

Gravitational radiation from a neutron star is commonly expected 
in its early stages of formation. In the approximatation of a 
neutron star of uniform mass density, rotation about centrifugal 
break-up gives the upper bound 
\begin{eqnarray}
\alpha^*=\frac{4}{15}M_{ns}R^2_{ns}\Omega_{ns}^3
=\frac{4}{15}(M_{ns}/R_{ns})^{5/2} 
\simeq 0.005
\label{EQN_T2}
\end{eqnarray} 
for a canonical value $M_{ns}/R_{ns}\simeq1/5$ of the mass-to-radius 
ratio of a neutron star. 
The true value of $\alpha^*$ may be somewhat less 
by an (uncertain) efficiency factor and extremal angular velocities 
below the Newtonian bound $M_{ns}^{1/2}/R_{ns}^{3/2}$.

\section{Calorimetric test for Kerr black holes}

Gravitational wave-emissions from a black hole-torus system
may be discriminated from those by a neutron star through the test
\begin{eqnarray}
\alpha>\alpha^*
\label{EQN_AB}
\end{eqnarray}
The estimates (\ref{EQN_T1},\ref{EQN_T2}) show
that (\ref{EQN_AB}) tends to be satisfied 
{\em whenever the black hole-torus 
system is more compact than a neutron star}. Here, compactness is 
now re-expressed in terms of a mass-to-radius ratio: 
the black hole mass to the radius of the torus in the former, and 
the neutron star mass to its radius in the latter case.

Applying the test (\ref{EQN_AB}) requires the distance 
to the source in deriving the fluence from the observed
strain amplitude. This may be circumvented using a sample
of detections, and taking averages to consider
\begin{eqnarray}
\bar{\alpha} >\alpha^*
\label{EQN_C}
\end{eqnarray}
in the approximation
\begin{eqnarray}
\bar{\alpha}\simeq 2\pi\bar{E}\bar{f}.
\label{EQN_AL}
\end{eqnarray}
A cosmologically nearby sample is well-described by 
a constant GRB event-rate $\dot{n}$ per unit volume,
e.g., 2 per year within a distance to 100Mpc.
This gives a differential event rate of gravitational wave-bursts
$d\dot{N}\simeq \Sigma \dot{n} dr,$ 
where $\Sigma=4\pi r^2$ denotes the surface area 
of the sphere reaching the source at radius $r$.
In what follows, we shall assume a source population
with standard $E$ and $f$. Consider a sample of  
surface energy densities $e=E/\Sigma$ at the detector.
Upon averaging, we have $\bar{e}=E<\Sigma_{max}/\Sigma>/\Sigma_{max}
 =3 E/\Sigma_{max}$, where $\Sigma_{max}=4\pi r_{max}$
 denotes the maximal surface area associated with a 
 source at $r=r_{max}$ at detection threshold.
The assumption that of standard emissions $E$ may be
checked through the relationship
\begin{eqnarray}
d\dot{N}\propto e^{-1/2} d(1/e).
\end{eqnarray}
The averaged expression (\ref{EQN_AL}) for $\alpha$ now becomes
\begin{eqnarray}
\bar{\alpha}=\left({4\pi}/{3}\right)^{5/3}
    \left({\dot{N}}/{\dot{n}}\right)^{2/3}
    \bar{e}\bar{f}
\end{eqnarray}
upon expressing $\Sigma_{max}$ in the net detection rate $\dot{N}$.

Sources of gravitational waves in cosmologically nearby sample
can be tested for an association with GBRs by comparing their durations with the 
redshift-corrected mean duration $\bar{T}=\mbox{30s}<1/(1+z)>$ of long GRBs in the 
BATSE catalogue \citep{kou93}. This gives $\bar{T}\simeq$10s if 
long GRBs are locked to the star formation rate, or $\bar{T}\simeq$10-15s working
from the measured redshifts $z$ to individual events (as in \cite{mvp01d,mvp01e}).
The latter might overestimate $\bar{T}$ due to extinction of afterglows 
(see \cite{ram01}). 
A typical redshift of about two would be consistent with 
$<V/V_{max}>=0.282$ for long bursts \citep{kat96} and the peak 
in the stellar formation rate (e.g., \cite{con97}).

In summary, calorimetry on gravitational wave-emissions 
from GRBs could offer a unique opportunity to establish the existence of 
Kerr black holes as objects in Nature, and hence that of black holes in general. 

{\bf Acknowledgement.} The author acknowledges stimulating discussions with
R.P. Kerr, G.J. Tee, T. Piran and G.E. Brown, and constructive comments from 
the anonymous referee.
This research is supported by NASA Grant 5-7012 and an MIT C.E. Reed Fund.

\end{document}